\documentclass[10pt,conference]{IEEEtran}

\usepackage{cite}
\usepackage{graphicx}
\usepackage{psfrag}
\usepackage{subfigure}
\usepackage{flushend}
\usepackage{color}

\usepackage{amsmath}
\usepackage{array}
\usepackage{graphics}
\usepackage{epsfig}
\usepackage{amsbsy}
\usepackage{amssymb}
\usepackage{amsthm}
\newtheorem{theorem}{Theorem}

\newtheorem{remark}[theorem]{Remark}

\usepackage{framed}
\usepackage{algorithm}
\newtheorem{proposition}{Proposition} 
\IEEEoverridecommandlockouts

\begin{document}

\title{\LARGE Outage Analysis of Spectrum Sharing Energy Harvesting\\ Cognitive Relays in Nakagami-$m$ Channels\vspace*{-2mm}}
\author{Sanket S.~Kalamkar\IEEEauthorrefmark{1}\IEEEauthorrefmark{3},
        Subhajit~Majhi\IEEEauthorrefmark{2},
        and~Adrish~Banerjee\IEEEauthorrefmark{1}\vspace*{-1mm}

\thanks{\IEEEauthorrefmark{1}The authors are with the Department of Electrical Engineering, Indian Institute of Technology, Kanpur, India (e-mail: kalamkar@iitk.ac.in, adrish@iitk.ac.in).}
\thanks{\IEEEauthorrefmark{2}The author is with the Department of Electrical and Computer Engineering, University of Waterloo, Waterloo, Canada (e-mail: smajhi@uwaterloo.ca).}
\thanks{\IEEEauthorrefmark{3}The author is supported by the TCS research scholarship.}

}

\maketitle

\begin{abstract}
Energy harvesting (EH) cognitive relays are an exciting solution to the problem of inefficient use of spectrum while achieving green communications and spatial diversity. In a spectrum sharing scenario, we investigate the performance of a cognitive relay network, where a secondary source communicates with its destination over Nakagami-$m$ channels via  decode-and-forward EH relays while maintaining the outage probability of the primary user below a predefined threshold. Specifically, we derive a closed-form expression for the secondary outage probability and show that it is a function of the probability of an EH relay having sufficient energy for relaying, which in turn, depends on the energy harvesting and consumption rates of the EH relay and the primary outage probability threshold. We also show that relaxing the primary outage constraint may not always benefit the cognitive EH relay network due to the limitations imposed on the relay's transmit power by the energy constraint. 
\end{abstract}\vspace*{-1mm}

\section{Introduction}\vspace*{-1mm}
 
With the ever-increasing demand for wireless services along with a need for green communications, spectral efficiency and energy efficiency have become important criteria in the design of future wireless systems. Energy harvesting (EH) cognitive radio \cite{sultan,park1,niyato2014,lee1,jeya} is a promising solution to improve the spectrum utilization; in particular, spectral efficiency is improved by spectrum sharing, while achieving self-sustaining green communications. In cognitive radio, a secondary user (SU) may share the spectrum with a primary user (PU) provided that the interference from it to PU remains below a given threshold \cite{luo}. 

The use of cooperative relays in cognitive radio has gained significant attention as they have the potential to improve the coverage and reliability of SU's transmission while sharing the spectrum with PU \cite{zhang1,guo,zou,si1,luo,duong1,tourki2012,zhong,lee,si10,weiw}. However, the relays may have limited battery reserves, and recharging or replacing the battery frequently may be inconvenient. This invokes the need for an external power source to keep relays active in the network. The EH relays can overcome such energy shortage while exploiting the spatial diversity~\cite{nasir,aissa,mehta2,krikidis,yener1}. As to the EH relays in cognitive radio, \cite{mousa,sanket} consider an EH secondary relay which helps relaying the secondary data, and perform the secondary outage analysis for Rayleigh fading channels under the interference constraint at the primary receiver; while in~\cite{van}, cooperative communication via multiple EH relays is considered. 

In this paper, we consider the case where SU uses the best relay from multiple EH relays for its own transmission over Nakagami-$m$ channels, given that PU's outage probability remains below a given threshold$-$we characterize the interference to PU by its outage probability. For EH relays, the optimal use of available energy is crucial. Low transmission power to conserve energy may prolong the lifetime of a relay, however, at the cost of increased outage; whereas higher transmission power improves the transmission quality, but at the expense of higher energy consumption rate reducing the future chances of transmission. Due to this EH nature of relays, the best relay selection becomes tricky as only relays having sufficient energy to forward the data to the destination, called \textit{active relays}, can be considered for the selection, making energy a crucial factor in the relay selection. Additionally, in spectrum sharing, the secondary communication via EH relays differs from that in non-spectrum sharing environment; because, in spectrum sharing, EH relay's transmit power depends not only on the energy availability with it, but also on the maximum power allowed by PU's outage constraint. For example, in a case where a relay has harvested less energy, it may not transmit with the maximum power allowed by PU's outage constraint. On the contrary, even if the relay has harvested large amount of energy, a tight PU's outage constraint may not allow relay to transmit with higher power. Thus, there exists an interesting tussle between these two constraints putting a stronger restriction on relay's transmit power than it would have been in the case with only one constraint, i.e., spectrum sharing without energy harvesting or non-spectrum sharing energy harvesting. Intrigued by the aforementioned tussle, in this work, we investigate its impact on the secondary network's performance, which is missing in~\cite{mousa,sanket,van}. The main contributions of this paper are as follows:
\begin{itemize}
\item Firstly, with the best relay selection scheme that maximizes SU's end-to-end signal-to-interference-noise-ratio (SINR), we specifically derive a closed-form expression for the outage probability of EH decode-and-forward (DF) cognitive relay network in Nakagami-$m$ channels under PU's outage constraint. We also consider the interference from PU while deriving the outage probability expression.
\item  Secondly, for better utilization of the harvested energy, we calculate the probability of a relay being active. We show that, besides energy harvesting and consumption rates, the probability of a relay being active depends on PU's outage probability threshold. We then couple the energy constraint due to the EH nature of relays with PU's outage constraint. We investigate which of the two constraints dominates the performance of EH relays and find the respective regions of dominance that regulate the transmit powers of EH relays.
\item Finally, we investigate the effects of fading severity parameter, number of relays, and the average energy harvesting rate on the secondary outage probability and tradeoff between the energy constraint and PU's outage constraint.
\end{itemize}
\vspace*{-2mm}

\section{Energy Harvesting and Spectrum Sharing Model}
\begin{figure}
\centering
\includegraphics[scale=0.2]{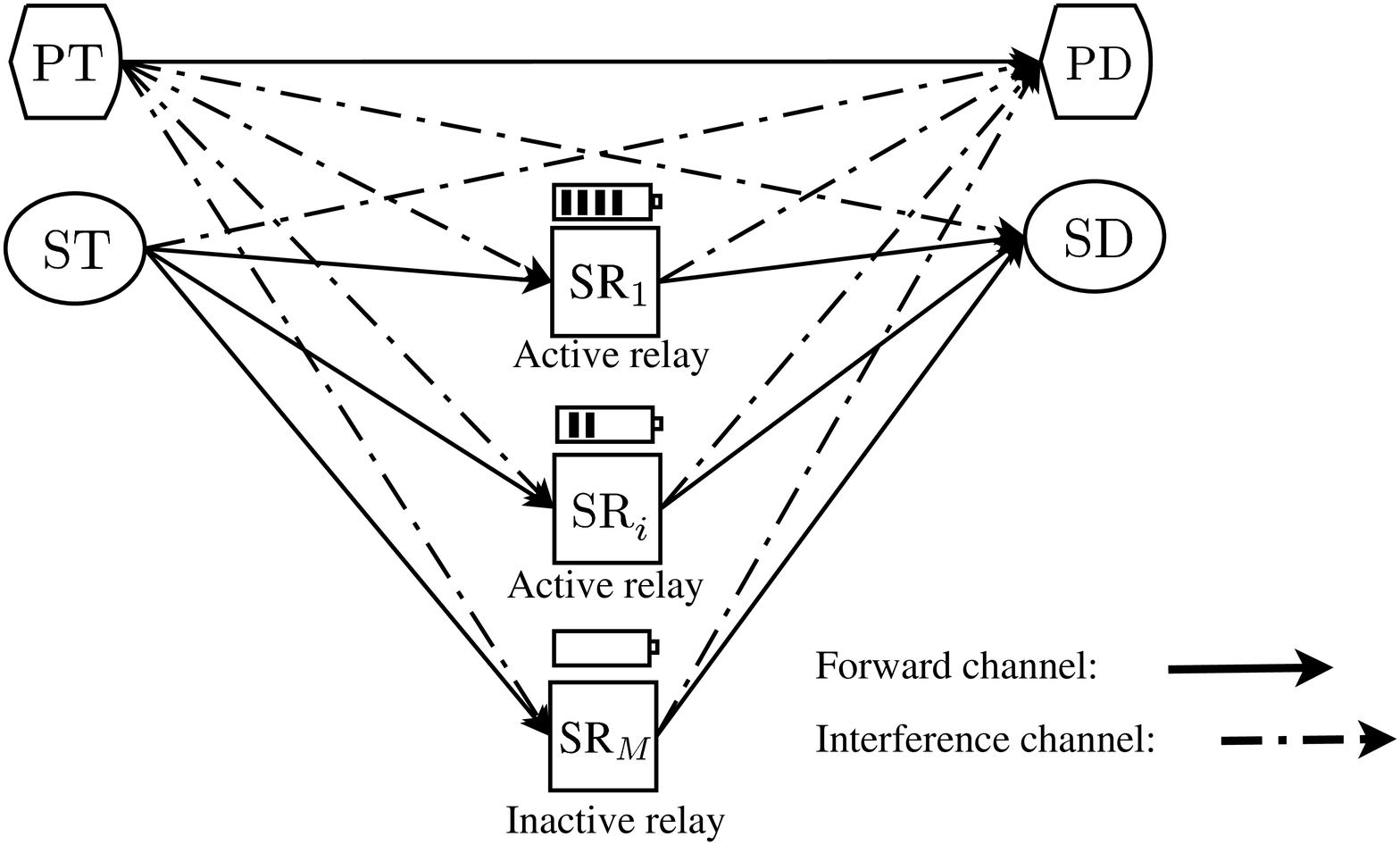}\vspace*{-3mm}
\caption{Secondary transmissions via EH relays with underlay spectrum sharing.}
\label{fig:syst}\vspace*{-6mm}
\end{figure}

\vspace*{-1mm}
As shown in Fig.\,\ref{fig:syst}, the network consists of a primary transmitter (PT), a primary destination (PD), a secondary transmitter (ST), a secondary destination (SD), and $M$ energy harvesting DF secondary relays (SRs). The PT, PD, ST, and SD are conventional nodes with constant energy supply (e.g., battery). 
The ST-SD direct link is assumed to be unavailable~\cite{zhong,lee,nasir,aissa,si10,weiw}. The ST communicates with SD over $i$th half-duplex EH relay ($\mathrm{SR}_i$), $i \in \lbrace1, 2, \dotsc, M \rbrace$. The channel between a transmitter $p \in \lbrace \mathrm{PT, ST, SR}_i \rbrace$ and a receiver $q \in \lbrace \mathrm{PD, SD, SR}_i\rbrace$, is a Nakagami-$m$ fading channel; $h_{p-q}$ denotes the channel coefficient. Thus, the channel power gain $|h_{p-q}|^2$ is Gamma-distributed with mean $\Omega_{p-q}$ and fading severity parameter $m_{p-q}$. We can write the probability density function and cumulative distribution function\,(CDF) of a random variable $U = |h_{p-q}|^2$ as\vspace*{-1mm}
\begin{eqnarray}
f_{U}(u) &=& \frac{\alpha_{p-q}^{m_{p-q}}}{\Gamma(m_{p-q})}u^{m_{p-q}-1}\exp(-\alpha_{p-q} u),
\label{eq:pdf}\vspace*{-1mm}
\end{eqnarray}\vspace*{-2mm}
\begin{eqnarray}
F_{U}(u) &=& \frac{\Upsilon(m_{p-q}, \alpha_{p-q} u)}{\Gamma(m_{p-q})} = 1 - \frac{\Gamma(m_{p-q}, \alpha_{p-q} u)}{\Gamma(m_{p-q})},
\label{eq:cdf}\vspace*{-3mm}
\end{eqnarray}\vspace*{-4mm}

\noindent respectively, where $\Gamma(\cdot)$, $\Gamma(\cdot, \cdot)$, and $\Upsilon(\cdot, \cdot)$ are the complete, upper incomplete, and lower incomplete Gamma functions \cite{gradshteyn}, respectively; $\alpha_{p-q} = m_{p-q}/\Omega_{p-q}$. The channels are independent of each other. For PT-PD, ST-PD, and SR$_{i}$-PD links, we assume the mean channel power gain knowledge due to limited feedback; while SR$_{i}$ and SD have the instantaneous channel gain knowledge for respective receiving links, i.e., ST-SR$_{i}$ and PT-SR$_i$ links at SR$_{i}$, and SR$_{i}$-SD and PT-SD links at SD~\cite{zou,peter:2013}. The secondary communication happens over two phases, each of $T$-second duration. All channels experience block-fading and remain constant for $\mathrm{2}T$-second, i.e., two phases of secondary communication, as in \cite{mehta2,aissa,nasir}. In phase $\mathrm{1}$, ST transmits to EH secondary relays, while in phase $\mathrm{2}$,  the received signal from ST is forwarded by one of the relays to SD after decoding. Note that in phase $\mathrm{2}$, no relay might be active due to the lack of energy required to forward the signal. 

\vspace*{-1mm}
\subsection{Energy Harvesting Model}\vspace*{-1mm}
The energy harvesting process of a relay $i$ is stationary and ergodic~\cite{mehta2}, with mean $H_{\mathrm{av}, i}$ Joules per second\,($\mathrm{J/s}$). This model encompasses different energy harvesting sources like solar, vibrations, radio frequency\,(RF) in the surroundings~\cite{lu}, and different energy harvesting profiles \cite{mehta1}. The EH relay stores the harvested energy in a battery with negligible leakage. For analytical tractability, we assume the capacity of the energy storage to be large\cite{aissa,mehta2,krikidis}. In addition, the energy consumption occurs only in data transmission; any other energy expenditure, e.g., energy consumption in signal reception and processing, is not considered for the purpose of exposition~\cite{mehta2,aissa,mousa}.\vspace*{-1mm}

\subsection{Maximum Secondary Transmit Powers in Spectrum Sharing}
In this work, we characterize the quality of service (QoS) of PU by its outage probability. For constant transmit power of PT ($P_{\mathrm{PT}}$), the PU outage probability should be below a certain threshold $\Theta_{\mathrm{p}}$ given the interference from the secondary transmitter and the relay. This constraint limits the transmit powers of ST and SR to $P_{\mathrm{ST}}$ and $P_{\mathrm{SR}}$, respectively.
In phase $\mathrm{1}$, the outage probability of PU $\mathrm{P_{p, out, ST}}$ when ST is transmitting, is given as\vspace*{-1mm}
\begin{align}
\mathrm{P_{p, out, ST}} = \mathrm{Pr}\left(\log_{2}\left(1+ \gamma_{\mathrm{PD}}\right)\leq\mathcal{R}_\mathrm{p}\right) \leq \Theta_{\mathrm{p}},\vspace*{-2mm}
\label{eq:3}\vspace*{-2mm}
\end{align}\vspace*{-4mm}

\noindent where $\gamma_{\mathrm{PD}} = \frac{P_{\mathrm{PT}}|h_{\mathrm{PT-PD}}|^2}{P_{\mathrm{ST}}|h_{\mathrm{ST-PD}}|^2 + N_0}$ is SINR at PD, $N_0$ being the noise power of additive white Gaussian noise (AWGN) at all receivers, and $\mathcal{R}_\mathrm{p}$ is the desired data rate on the primary link.\vspace*{-2mm}
\begin{proposition}
 We write $\mathrm{P_{p, out, ST}}$ as follows:

\vspace*{-4mm}
{{\small
\begin{align}
\mathrm{P_{p, out, ST}}&= 1-\Bigg[\frac{\alpha_{\mathrm{ST-PD}}^{m_{\mathrm{ST-PD}}}\exp\left(\frac{-\alpha_{\mathrm{PT-PD}}\theta_\mathrm{p}N_0}{P_{\mathrm{PT}}}\right)}{\Gamma(m_{\mathrm{ST-PD}})} \nonumber \\
&\times\sum_{k=0}^{m_{\mathrm{PT-PD}}-1}\left(\frac{\alpha_{\mathrm{PT-PD}}\theta_\mathrm{p}N_0}{P_{\mathrm{PT}}}\right)^k\frac{1}{k!} \sum_{t=0}^{k}{\binom{k}{t}}\left(\frac{P_{\mathrm{ST}}}{N_0}\right)^t \nonumber \\
& \times\frac{\Gamma(m_{\mathrm{ST-PD}} + t)}{\left(\frac{\alpha_{\mathrm{PT-PD}}\theta_\mathrm{p} P_{\mathrm{ST}}}{P_{\mathrm{PT}}} + \alpha_{\mathrm{ST-PD}}\right)^{m_{\mathrm{ST-PD}} + t}}\Bigg].
\label{eq:power}
\end{align}}}
\end{proposition}\vspace*{-1mm}
\begin{proof}
The proof is given in Appendix \ref{sec:der_out}.
\end{proof}
Using \eqref{eq:power}, the value of maximum ST power $P_{\mathrm{ST}}$ allowed by PU's outage constraint $\Theta_{\mathrm{p}}$ can be numerically found. Similarly, in phase $\mathrm{2}$, the maximum allowable transmit power $P_{\mathrm{SR}}$ for a relay can be numerically found from \eqref{eq:power1}, replacing the role of secondary transmitter in \eqref{eq:power} by the secondary relay and replacing corresponding channel parameters.\vspace*{-4mm}

{{\small
\begin{align}
\mathrm{P_{p, out, SR}} &= 1-\Bigg[\frac{\alpha_{\mathrm{SR-PD}}^{m_{\mathrm{SR-PD}}}\exp\left(\frac{-\alpha_{\mathrm{PT-PD}}\theta_\mathrm{p}N_0}{P_{\mathrm{PT}}}\right)}{\Gamma(m_{\mathrm{SR-PD}})}\nonumber \\
& \times \sum_{k=0}^{m_{\mathrm{PT-PD}}-1}\left(\frac{\alpha_{\mathrm{PT-PD}}\theta_\mathrm{p}N_0}{P_{\mathrm{PT}}}\right)^k\frac{1}{k!}  \sum_{t=0}^{k}{\binom{k}{t}}\left(\frac{P_{\mathrm{SR}}}{N_0}\right)^t \nonumber \\
& \times \frac{\Gamma(m_{\mathrm{SR-PD}} + t)}{\left(\frac{\alpha_{\mathrm{PT-PD}}\theta_\mathrm{p} P_{\mathrm{SR}}}{P_{\mathrm{PT}}} + \alpha_{\mathrm{SR-PD}}\right)^{m_{\mathrm{SR-PD}} + t}}\Bigg].
\label{eq:power1}
\end{align}}}\vspace*{-4mm}

\subsection{Active Relays and Best Relay Selection}\vspace*{-1mm}
Assume that out of the total $M$ relays, $N$ relays are active $\left(N \in \lbrace 0, 1, \dotsc, M\rbrace\right)$ due to energy availability, and a relay has to be selected from active $N$ relays. An \textit{active} relay is the relay having sufficient energy to forward the received data from ST.

For an opportunistic DF relaying, the relay with the largest end-to-end SINR at SD, called \textit{the best relay}, is selected to forward the signal. When $N$ relays are available for selection, the largest end-to-end SINR at SD is given by\vspace*{-1mm}
\begin{equation}
\gamma_{\mathrm{tot}}^{N}=  \max_{\mathrm{SR}_i \in \mathbb{R}}(\min(\gamma_{\mathrm{SR}_i}, \gamma_{\mathrm{R}_i\mathrm{D}})),
\label{eq:rn1}\vspace*{-1mm}
\end{equation}
where $\mathbb{R}$ is the set of active relays. Note that $\mathbb{R}$ is an empty set when no relay is active. $\gamma_{\mathrm{SR}_i}$ and $\gamma_{\mathrm{R}_i\mathrm{D}}$ are SINRs at $i$th relay and at SD over $\mathrm{R}_i-\mathrm{D}$ channel, respectively, and are given as\vspace*{-2mm}
\begin{equation}
\gamma_{\mathrm{SR}_i} =  \frac{P_{\mathrm{ST}}|h_{\mathrm{ST}-\mathrm{SR}_i}|^2}{P_{\mathrm{PT}}|h_{\mathrm{PT}-\mathrm{SR}_i}|^2 + N_0},
\label{eq:pst11}\vspace*{-1mm}
\end{equation}
\begin{equation}
\gamma_{\mathrm{R}_i\mathrm{D}} =  \frac{P_{\mathrm{SR}_i}|h_{\mathrm{SR}_i-\mathrm{SD}}|^2}{P_{\mathrm{PT}}|h_{\mathrm{PT-SD}}|^2 + N_0},\vspace*{-1mm}
\label{eq:psr11}
\end{equation}
where we obtain $P_{\mathrm{ST}}$ and $P_{\mathrm{SR}_i}$ from \eqref{eq:power} and \eqref{eq:power1}, respectively. 
\section{Secondary Outage Analysis}
Now, when we select the best relay out of $N$ active relays, the secondary outage probability $\mathrm{P}^{N}_{\mathrm{s,out}}$ can be given as\vspace*{-2mm}
\begin{equation}
\mathrm{P}^{N}_{\mathrm{s,out}}(\gamma) \!=\! 
\mathrm{Pr}(\gamma_{\mathrm{tot}}^{N} \leq \gamma) \!=\! \mathrm{Pr}\!\left(\!\max_{\mathrm{SR}_i \in \mathbb{R}}(\min(\gamma_{\mathrm{SR}_i}, \gamma_{\mathrm{R}_i\mathrm{D}})) \!\leq\! \gamma\!\right)\!,
\label{eq:psout_basic}\vspace*{-1mm}
\end{equation}
where secondary's desired secondary rate is $\mathcal{R}_{\mathrm{s}} = \frac{1}{2}\log_2\left(1 + \gamma\right)$. For the ease of representation and without compromising the insight into analysis, we consider $m_{\mathrm{ST}-\mathrm{SR}_i}\!=\!m_{\mathrm{ST-SR}}$, $m_{\mathrm{SR}_i-\mathrm{SD}} \!=\! m_{\mathrm{SR-SD}}$, $m_{\mathrm{PT}-\mathrm{SR}_i}\!=\! m_{\mathrm{PT-SR}}$, $m_{\mathrm{SR}_i-\mathrm{PD}}\! = \!m_{\mathrm{SR-PD}}$, $\Omega_{\mathrm{ST}-\mathrm{SR}_i} \!=\! \Omega_{\mathrm{ST-SR}}$, $\Omega_{\mathrm{SR}_i-\mathrm{SD}} \!= \!\Omega_{\mathrm{SR-SD}}$, $\Omega_{\mathrm{PT}-\mathrm{SR}_i} \!=\! \Omega_{\mathrm{PT-SR}}$, and $\Omega_{\mathrm{SR}_i-\mathrm{PD}}\! =\! \Omega_{\mathrm{SR-PD}}$. Then, $P_{\mathrm{SR}_i} = P_\mathrm{SR}$. Below we give the closed-form expression for $\mathrm{P}^{N}_{\mathrm{s,out}}$.
\begin{proposition}
We write $\mathrm{P}^{N}_{\mathrm{s,out}}(\gamma)$ as follows:\vspace*{-2mm}
\begin{align}
{\mathrm{P}}^{N}_{\mathrm{s, out}}(\gamma) &=\sum_{r_0=0}^{N}\sum_{r_1=0}^{r_0}\!\dotsc\!\!\!\!\!\! \sum_{r_{m_{\mathrm{SR-SD}}-1}=0}^{r_{m_{\mathrm{SR-SD}}-2}}\!\!{\binom{N}{r_0}}\!\!{\binom{r_0}{r_1}}\!\dotsc\!{\binom{r_{m_{\mathrm{SR-SD}}-2}}{r_{m_{\mathrm{SR-SD}}-1}}}\nonumber\\
&\!\!\times\mathcal{A}^{N-r_{m_\mathrm{SR-SD}-1}}	(-1)^{N+r_{m_{\mathrm{SR-SD}}-1}}  \nonumber \\ 
&\times  \Bigg[\mathrm{exp}\left( -\frac{\alpha_{\mathrm{SR-SD}}\gamma N_0(N-r_{m_\mathrm{SR-SD}-1})}{P_{\mathrm{SR}}} \right) \nonumber\\
&\times\left(\frac{\alpha_{\mathrm{SR-SD}}\gamma N_0}{P_{\mathrm{SR}}}\right)^{R_{m_\mathrm{SR-SD}}}  \prod_{k=1}^{m_\mathrm{SR-SD}-1} \left( \frac{1}{k!}\right)^{r_{k-1}- r_{k}} \Bigg] \nonumber \\
&\times  \Bigg[\sum_{p =0}^{R}{\binom{R}{p}}\left(\frac{P_{\mathrm{PT}}}{N_0}\right)^p\frac{1}{(m_{\mathrm{PT-SD}}-1)!}\nonumber\\
&\!\times\!\! \frac{\alpha_{\mathrm{PT-SD}}^{m_{\mathrm{PT-SD}}}(m_{\mathrm{PT-SD}}+p-1)!}{\!\left(\!\alpha_{\mathrm{PT-SD}}+\frac{\alpha_{\mathrm{SR-SD}}\gamma P_{\mathrm{PT}} (N-r_{m_{\mathrm{SR-SD}} -1})}{P_{\mathrm{SR}}}\!\right)^{m_{\mathrm{PT-SD}}+p}}\Bigg],
\label{eq:psout}
\end{align}\vspace*{-5mm}

\noindent where $\mathcal{A}$ is given by \eqref{eq:A}.
\end{proposition}\vspace*{-2mm}
\begin{proof}
See Appendix \ref{appen:2}. A key idea in the proof is to consider the dependency between $\gamma_{\mathrm{R}_i\mathrm{D}}$ and $\gamma_{\mathrm{R}_j\mathrm{D}}$ $(i \neq j, j \in \lbrace 1, 2, \dotsc, M\rbrace)$, originating due to the common term $|h_{\mathrm{PT-SD}}|^2$.
\end{proof}

For an EH relay, its operation is subject to the energy neutrality constraint, which states that a relay cannot spend more energy than it has harvested. Thus, it is possible that the relay might remain inactive for some time due to the lack of energy.

Let us denote the probability of a relay $i$ being active by $\eta_i \geq 0$. In a non-spectrum sharing scenario, $\eta_i$ depends on the relay's energy harvesting and consumption rates.  Based on these two factors, the relay $i$ operates in two regions as follows:
\begin{itemize}
\item \textit{Energy constrained region ($\eta_i < 1$})
\item \textit{Energy unconstrained region ($\eta_i = 1$)}.
\end{itemize}
A relay operates in the energy unconstrained region if its average energy consumption rate is less than the average energy harvesting rate, i.e., the relay is always active.

We assume $H_{\mathrm{av},i} = H_{\mathrm{av}} $ without loss of generality. Then, we have $\eta_i = \eta$. The energy available with the relay depends on the factors that, how frequently the relay is selected; its harvested energy till now; and when was the energy harvested. As we will show later, in the case of spectrum sharing with PU, the probability of a relay being active, i.e., $\eta$, depends not only on the energy harvesting rate, the total number of relays in the system, the energy consumed by a relay in its each transmission, but also on PU's outage constraint. Using the following proposition given in~\cite{mehta2}, we show the dependency of $\eta$ on PU's outage constraint.\vspace*{-2mm}

\begin{proposition}
\label{prop:main}
Let the probability of selecting a relay be $\omega$. Then, $\omega = \frac{2H_{\mathrm{av}}}{P_{\mathrm{SR}}}$. The relays remain active with the probability\vspace*{-1mm}
\begin{equation}
\eta = 1 - \left[(1 - M\omega)^{+}\right]^{\frac{1}{M}}.\vspace*{-1mm}
\label{eq:rel}\vspace*{-1mm}
\end{equation}
All the relays become energy unconstrained, i.e., $\eta = 1$, when $\omega \geq 1/M$. We denote $(x)^{+} = \max(0, x)$.
\end{proposition}\vspace*{-2mm}

The expression for $\omega$ in Proposition $\mathrm{3}$ is obtained from the energy neutrality constraint and stationarity and ergodicity of the energy harvesting process. From Proposition~\ref{prop:main}, one can notice that the probability of a relay being active depends on the power $P_{\mathrm{SR}}$ with which the relay performs a transmission. Equations \eqref{eq:power1} and \eqref{eq:rel} together show that $P_{\mathrm{SR}}$, in turn, the probability of a relay being active, depends on the primary outage constraint.

Given $N$ out of $M$ relays are active, each with the probability $\eta$, we obtain the final expression for the secondary outage probability with EH relays by unconditioning over the number of active relays as\vspace*{-1mm}
\begin{equation}
\mathrm{P_{s,out}} = \sum_{N = 1}^{M} {\binom{M}{N}} \eta^{N}(1 - \eta)^{M-N}{\mathrm{P}}^{N}_{\mathrm{s, out}} + (1-\eta)^{M}\mathrm{P_{s,out}^0},\vspace*{-2mm}
\label{eq:finale}
\end{equation}\vspace*{-1mm}

\noindent where ${\mathrm{P}}^{N}_{\mathrm{s, out}}$ given by \eqref{eq:psout} is the secondary outage probability when we select the best relay among active $N$ relays to forward the signal from ST; ${\mathrm{P}}^{0}_{\mathrm{s, out}}$ is the secondary outage probability when no relay is active, and is equal to 1.

\section{Discussions and Results}
In spectrum sharing, PU's outage constraint governs the maximum transmit power of relays. In addition, if relays are energy harvesting, due to the limited available harvested energy, the probability of a relay being active plays an important role in the performance of the secondary system. In this section, we will first discuss the effect of PU's outage constraint on SU's outage performance for the case when an EH relay on selection, uses the maximum power allowed by the primary, aiming to reduce the secondary outage probability; even though, it might also reduce the relay's probability of being active. Then, we will consider the case when, along with PU's outage constraint, EH relays aim to keep their probability of being active to one ($\eta = 1$)$-$which we will call the \textit{energy constraint}{\footnote{Note that the constraint $\eta = 1$ is different from the \textit{energy neutrality constraint}. With latter, the consumed energy by a relay cannot exceed its harvested energy, whereas the constraint of always being active does not allow relay's energy consumption rate to exceed its energy harvesting rate. With the higher energy consumption rate, the relays will eventually consume the energy harvested in the past before acquiring sufficient newly harvested energy to keep them active.}}. In this case, we will show that the transmit power of the relay is regulated by the dominant of the two constraints$-$PU's outage constraint and energy constraint$-$and we will find the region of dominance for each constraint. Finally, we will see from results that in both the above cases, relaxing PU's outage constraint beyond a level does not offer any benefit to the secondary system with EH relays.
\subsection{System Parameters and Simulation Setup}
We consider following parameter values: PU transmit power, $P_\mathrm{PT} = \mathrm{15}$\,$\mathrm{dB}$; the desired primary rate, $\mathcal{R}_{\mathrm{p}}$  = $\mathrm{0.4}$\,$\mathrm{bits/s/Hz}$; the desired secondary rate, $\mathcal{R}_{\mathrm{s}}$  = $\mathrm{0.2}$\,$\mathrm{bits/s/Hz}$; noise power, $N_0$ = $-\mathrm{60}~\mathrm{dBm}$. Denote fading severity parameters on forward and interference channels by $m_{\mathrm{f}}$ and $m_{\mathrm{int}}$, respectively. We consider a 2-D topology, where ($x_{i}$, $y_{i}$) defines the coordinate of $i$th user. The mean channel gain between $i$th user with coordinate ($x_{i}$, $y_{i}$) and $j$th user with coordinate ($x_{j}$, $y_{j}$) is $d_{ij}^{-\Delta}$, where $d_{ij}$ is the distance between users $i$ and $j$ in meters and $\Delta$ is the path-loss coefficient which is assumed to be $\mathrm{4}$. Without any loss of generality, ST is placed at ($\mathrm{0}$, $\mathrm{0}$), $M$ relays are clustered and collocated at ($\mathrm{50}$, $\mathrm{0}$), and SD is placed at ($\mathrm{100}$, $\mathrm{0}$). Also, PT and PD are located at ($\mathrm{50}$, $\mathrm{50}$) and ($\mathrm{100}$, $\mathrm{50}$), respectively.\vspace*{-1mm}

\subsection{Effect of PU's Outage Constraint}
\label{sec22}
\begin{figure}
\centering
\includegraphics[scale=0.39]{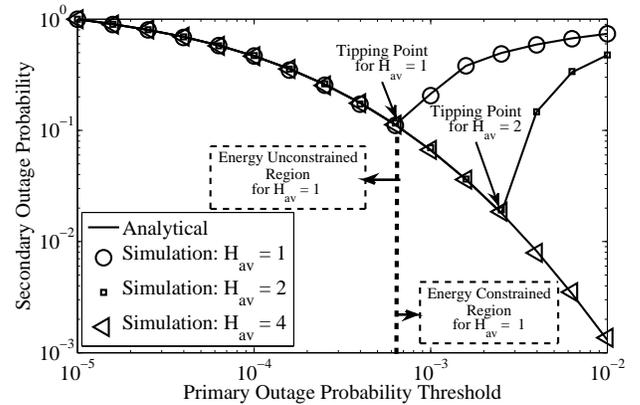}\vspace*{-3mm}
\caption{Secondary outage probability ($P_{\mathrm{s, out}}$) vs. primary outage probability threshold ($\Theta_{\mathrm{p}}$), $M = 3$, $m_{\mathrm{f}} = \mathrm{2}$, $m_{\mathrm{int}} = \mathrm{1}$.}
\label{fig:11}\vspace*{-4mm}
\end{figure}

\begin{figure}
\centering
\includegraphics[scale=0.39]{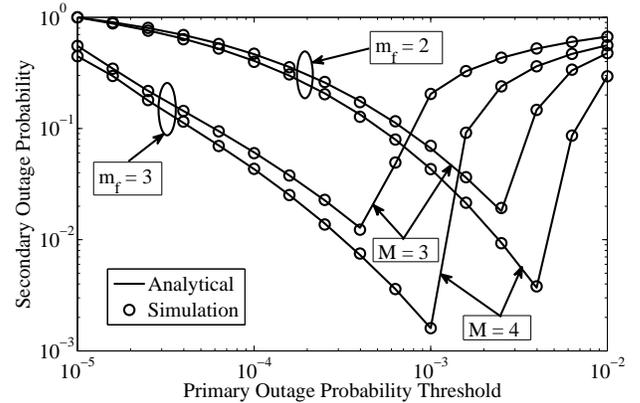}\vspace*{-3mm}
\caption{Secondary outage probability ($P_{\mathrm{s, out}}$) vs. primary outage probability threshold ($\Theta_{\mathrm{p}}$), effect of fading severity parameter and number of relays $M$, $m_{\mathrm{int}} = \mathrm{1}$, $H_{\mathrm{av}} = \mathrm{2}$\,$\mathrm{J/s}$.}
\label{fig:22}\vspace*{-5mm}
\end{figure}

Figs.\,\ref{fig:11} and \ref{fig:22} show the effect of PU's outage constraint on the outage probability of the secondary system with EH relays.\footnote{Simulation results validate the analysis. The number of iterations is up to $\mathrm{10^6}$.} The selected EH relay transmits with the maximum power allowed by PU's outage constraint. We notice that the increase in the primary outage threshold $\Theta_{\mathrm{p}}$ increases the maximum transmit power $P_{\mathrm{SR}}$ allowed for the relay, which initially reduces the secondary outage probability $P_{\mathrm{s, out}}$. However, with the increase in the threshold $\Theta_{\mathrm{p}}$ beyond a level, a \textit{tipping point} will be reached after which $P_{\mathrm{s, out}}$ will increase even with the increase in $P_{\mathrm{SR}}$ as relays will consume energy at a higher rate than they will harvest, i.e., relays will become \textit{energy constrained} (see plots for $H_\mathrm{av}$\,=\,$\mathrm{1}, \mathrm{2}$ in Fig.\,\ref{fig:11}). This will reduce the probability of a relay being active, thereby reducing the number of relays available to forward the data to SD. As long as $P_{\mathrm{SR}}$ is below a certain level so that $\omega$\,=\,$\frac{2H_{\mathrm{av}}}{P_{\mathrm{SR}}}$\,$\geq$\,$1/M$ as shown in Proposition $\mathrm{3}$, the relays operate in the \textit{energy unconstrained region}, i.e., the harvested power is more than the transmit power $P_{\mathrm{SR}}$. But, with relaxation of PU's outage constraint, eventually the value of $P_{\mathrm{SR}}$ increases such that $\omega$\,$<$\,$1/M$, making relays energy constrained and increasing $P_{\mathrm{s, out}}$. Also, increase in the harvesting rate $H_{\mathrm{av}}$ delays the occurrence of the \textit{tipping point} as expected, and at high harvesting rates, the relays might operate completely in the energy unconstrained region due to availability of abundant energy (see the plot for $H_\mathrm{av}$\,=\,$\mathrm{4}$ in Fig.\,\ref{fig:11}).\vspace*{-1mm} 
\begin{remark}
Relaxing PU's outage constraint may not always improve the performance of EH secondary relays in spectrum sharing. That is, unlike conventional non-EH case, due to lack of energy, the relays with EH capability may not transmit with the maximum allowed power even though they are allowed to do so.
\end{remark}\vspace*{-3mm}

\subsection{Effect of Fading Severity Parameter}
Fig.\,\ref{fig:22} shows the effect of fading severity parameter on $P_{\mathrm{s, out}}$. We notice that, before the \textit{tipping point}, i.e, in the energy unconstrained region, $P_{\mathrm{s, out}}$ is lower for higher fading severity parameter $m_{\mathrm{f}}$ on forward channels. However, the trend reverses after the \textit{tipping point}. This is because, with the increase in $m_{\mathrm{f}}$, the fading effect subsides over the primary link between PT and PD, providing an extra margin for maximum secondary relay transmit  power $P_{\mathrm{SR}}$ for a given $\Theta_{\mathrm{p}}$. This helps in achieving lower $P_{\mathrm{s, out}}$ for higher $m_{\mathrm{f}}$ in the energy unconstrained region, where the energy harvesting rate is higher than the energy consumption rate. As shown in Fig.\,\ref{fig:22}, for a given harvesting rate, due to higher allowed $P_{\mathrm{SR}}$ (higher energy consumption rate) for higher $m_{\mathrm{f}}$, the \textit{tipping point} arrives earlier than that for lower $m_{\mathrm{f}}$. After the \textit{tipping point}, since relays enter the energy constrained region, higher $m_{\mathrm{f}}$, in turn, higher energy consumption rate, reduces the probability of a relay being active. This often leads to non-availability of relays for transmission, increasing $P_{\mathrm{s, out}}$ for higher $m_{\mathrm{f}}$. Also, increase in the number of relays $M$ increases the probability of being active (see \eqref{eq:rel}) as the candidate relays for cooperation increases (increased diversity) due to which a certain relay is chosen less frequently. This reduces $P_{\mathrm{s, out}}$.
\begin{figure}
\centering
\includegraphics[scale=0.39]{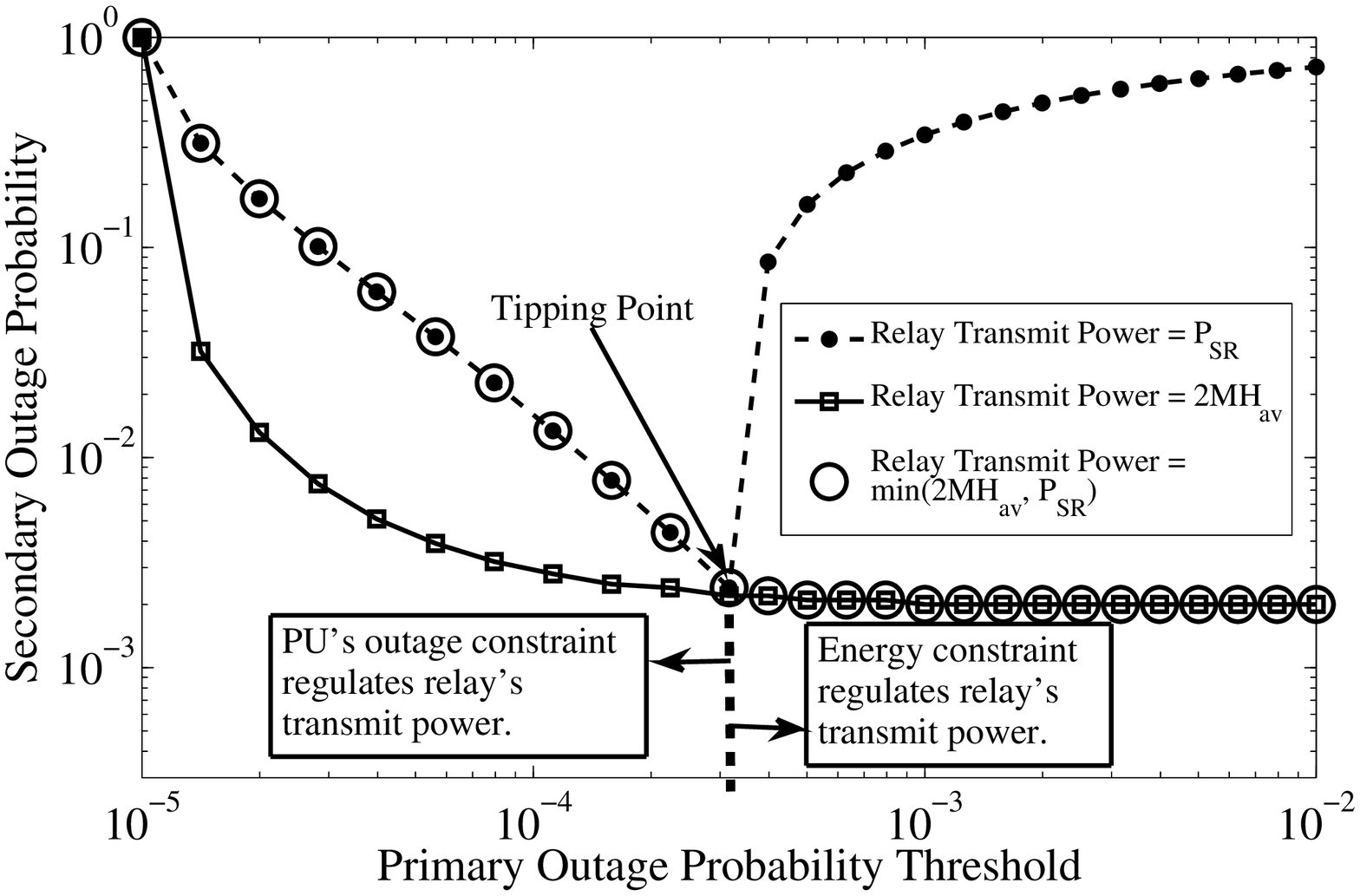}\vspace*{-3mm}
\caption{Secondary outage probability ($P_{\mathrm{s, out}}$) vs. primary outage probability threshold ($\Theta_{\mathrm{p}}$),  $M$ = $\mathrm{3}$, $m_{\mathrm{f}}= \mathrm{3}$, $m_{\mathrm{int}} = \mathrm{2}$, $H_{\mathrm{av}} = \mathrm{2}$\,$\mathrm{J/s}$.}
\label{fig:33}\vspace*{-7mm}
\end{figure}
\subsection{Joint Effect of PU's Outage Constraint and Energy Constraint}
From discussions of Figs.\,\ref{fig:11} and \ref{fig:22}, we note that EH relays being inconsiderate towards their probability of being active, leads to SU's inferior outage performance beyond the tipping point. Now, for instance, assume that EH relays try to remain always active ($\eta = \mathrm{1}$), i.e., try to satisfy the energy constraint, irrespective of PU's outage constraint, and transmit with power $P_{\mathrm{s,a}}$. From Proposition $\mathrm{3}$, we can see that satisfying the energy constraint corresponds to $\omega$\,$\geq$\,$1/M$, i.e., $P_{\mathrm{s,a}}$\,$\leq$\,$2MH_{\mathrm{av}}$. That is, as long as EH relays transmit with power no greater than $2MH_{\mathrm{av}}$, they always remain active. Now, if we combine the energy constraint with PU's outage constraint, Fig.\,\ref{fig:33} shows that in the \textit{energy unconstrained region}, though an EH relay may transmit with maximum power $2MH_{\mathrm{av}}$ maintaining $\eta = \mathrm{1}$, the power $2MH_{\mathrm{av}}$ does not satisfy PU's outage constraint, i.e., $2MH_{\mathrm{av}} > P_{\mathrm{SR}}$. This leads to higher $P_{\mathrm{s, out}}$ in EH relay spectrum sharing scenario governed by both the energy constraint and PU's constraint than it would have been in EH non-spectrum sharing scenario governed by the energy constraint alone. In the \textit{energy constrained region}, the energy constraint becomes dominant, i.e., $2MH_{\mathrm{av}}$\,$<$\,$P_{\mathrm{SR}}$. Thus, even though PU's outage constraint is satisfied, and allows EH relays to transmit with the maximum power $P_{\mathrm{SR}}$, the energy constraint is violated, causing $\eta < \mathrm{1}$ and increasing in $P_{\mathrm{s, out}}$ as discussed for Figs.\,\ref{fig:11} and \ref{fig:22}. Therefore, we can see from the above discussion that, to satisfy both constraints, the maximum power with which EH relays may transmit is $\min(2MH_\mathrm{av}, P_\mathrm{SR})$. As shown in Fig.\,\ref{fig:33}, in the energy constrained region (after the tipping point), though transmitting with power $\min(2MH_\mathrm{av}, P_\mathrm{SR})$ avoids the increase in $P_{\mathrm{s, out}}$, relaxing PU's outage constraint does not improve SU's outage performance.\vspace*{-1mm}

\section{Conclusions}\vspace*{-1mm}
Under the primary outage constraint, this paper has analyzed the outage performance of the secondary communication via energy harvesting relays in Nakagami-$m$ channel. In a spectrum sharing scenario, the results show that, besides energy harvesting nature of relays, the primary outage constraint also strongly influences the probability of a relay being active. We note that relays should keep their probability of being active to one; otherwise, obeying only the primary outage constraint may lead to the inferior secondary outage performance. That is, in energy harvesting spectrum sharing, due to the energy constraint, relaxing the primary outage constraint may not always improve secondary outage performance unlike in non-energy harvesting case. Further, we have found the region of dominance for each of the constraints and proposed the optimal transmit power for relays to subside their inferior performance in the energy constrained region. We observe that increase in the number relays and lower fading severity parameter delays the entry of relays into the energy constrained region, boosting the secondary outage performance.\vspace*{-1mm}

\appendices
\section{Proof of \eqref{eq:power}}
\label{sec:der_out}
From \eqref{eq:3}, conditioned on $|h_{\mathrm{ST-PD}}|^2 = x$ we can write \vspace*{-5mm}

{{\small\begin{equation}
\mathrm{P_{p, out, ST}}\bigg|_{|h_{\mathrm{ST-PD}}|^2 = x}\!\!= \frac{\Upsilon\bigg(\!\!m_{\mathrm{PT-PD}}, \alpha_{\mathrm{PT-PD}} \frac{\theta_\mathrm{p} (P_{\mathrm{ST}}x + N_0)}{P_{\mathrm{PT}}}\!\!\bigg)}{\Gamma(m_{\mathrm{PT-PD}})}.
\label{eq:out1}\vspace*{-4mm}
\end{equation}}}

\noindent When $m_{\mathrm{PT-PD}}$ is a positive integer, we can write the lower incomplete Gamma function as \cite[8.352]{gradshteyn}\vspace*{-3mm}

{{\small\begin{equation}
\Upsilon(a, b) = (a-1)!\left(1-\exp(-b)\sum_{k=0}^{a-1}\frac{b^k}{k!}\right).
\label{eq:finite}\vspace*{-4mm}
\end{equation}}}

\noindent Then, using \eqref{eq:finite} in \eqref{eq:out1} and unconditioning over $|h_{\mathrm{ST-PD}}|^2$, we can write \eqref{eq:3} as\vspace*{-1mm}
%
\begin{eqnarray}
\mathrm{P_{p, out, ST}} \!\!\!\!\!\!&=&\!\!\!\! \int_0^{\infty}\mathrm{P_{p, out}}\bigg|_{|h_{\mathrm{ST-PD}}|^2 = x}\frac{\alpha_{\mathrm{ST-PD}}^{m_{\mathrm{ST-PD}}}}{\Gamma(m_{\mathrm{ST-PD}})}x^{m_{\mathrm{ST-PD}}-1}\nonumber\\
&& \times \exp(-\alpha_{\mathrm{ST-PD}} x)\mathrm{d}x.\vspace*{-2mm}
\label{eq:rand1}
\end{eqnarray}\vspace*{-6mm}

\noindent Simplifying \eqref{eq:rand1} and using binomial expansion, we get\vspace*{-4mm}

{{\small
\begin{align}
\mathrm{P_{p, out, ST}} &= 1-\frac{\alpha_{\mathrm{ST-PD}}^{m_{\mathrm{ST-PD}}}\exp\left(\frac{-\alpha_{\mathrm{PT-PD}}\theta_\mathrm{p}N_0}{P_{\mathrm{PT}}}\right)}{\Gamma(m_{\mathrm{ST-PD}})} \nonumber \\
&\!\!\times\! \sum_{k=0}^{m_{\mathrm{PT-PD}}-1}\left(\frac{\alpha_{\mathrm{PT-PD}}\theta_\mathrm{p}N_0}{P_{\mathrm{PT}}}\right)^k\frac{1}{k!} \sum_{t=0}^{k}{\binom{k}{t}}\left(\frac{P_{\mathrm{ST}}}{N_0}\right)^t \nonumber \\
&\!\!\times\!\! \int_{0}^{\infty}\!\!\!x^{m_{\mathrm{ST-PD}}+t-1}\!\exp\!\left(\!\!-x\!\left(\!\!\frac{\alpha_{\mathrm{PT-PD}}\theta_\mathrm{p} P_{\mathrm{ST}}}{P_{\mathrm{PT}}} \!+\! \alpha_{\mathrm{ST-PD}}\!\!\right)\!\right)\!\mathrm{d}x.
\label{eq:long1}
\end{align}}}\vspace*{-3mm}

\noindent Solving \eqref{eq:long1}, we get the required expression in \eqref{eq:power}.\vspace*{-1mm}

%

\section{Proof of \eqref{eq:psout}}
\label{appen:2}

From\,\eqref{eq:psr11}, we can see that $\gamma_{\mathrm{R}_i\mathrm{D}}$ and $\gamma_{\mathrm{R}_j\mathrm{D}}$ $(i \neq j, j \in \lbrace 1, 2, \dotsc, M\rbrace)$ contain the common term $|h_{\mathrm{PT-SD}}|^2$, which makes them dependent. Thus, conditioned on $|h_{\mathrm{PT-SD}}|^2 = x$, we can write the CDF of $\gamma_{\mathrm{tot}}^{N}$ as
\begin{eqnarray}
\mathrm{Pr}\left(\gamma_{\mathrm{tot}}^{N} \leq \gamma \big|_{|h_{\mathrm{PT-SD}}|^2 = x}\right) \!\!\!\!&=&\!\!\!\! \prod_{i = 1}^{N} \big[1 - \left(1 -\mathrm{Pr}\left(\gamma_{\mathrm{SR}_i} \leq \gamma \right) \right) \nonumber \\
&& \hspace*{-20mm}\times \underbrace{\left(1 - \mathrm{Pr} \left(\gamma_{\mathrm{R}_i\mathrm{D}} \leq \gamma\big|_{|h_{\mathrm{PT-SD}}|^2 = x} \right)\right)}_{\mathcal{I}}\bigg]. \vspace*{-3mm}
\label{eq:main}
\end{eqnarray}\vspace*{-4mm}
%

\noindent Using \eqref{eq:cdf}, we can write $\mathcal{I}$ in \eqref{eq:main} as\vspace*{-2mm}
\begin{eqnarray}
\mathcal{I}&=& \frac{\Gamma\left(m_{\mathrm{SR-SD}}, \alpha_{\mathrm{SR-SD}} \frac{\gamma(P_{\mathrm{PT}}x + N_0)}{P_{\mathrm{SR}}}\right)}{\Gamma\left(m_{\mathrm{SR-SD}}\right)}.
\label{eq:I}\vspace*{-1mm}
\end{eqnarray}
Using \cite[8.352]{gradshteyn}\vspace*{-1mm}
\begin{equation}
\Gamma(k, t) = (k-1)!\exp(-t)\sum_{n=0}^{k-1}\dfrac{t^n}{n!}, k=1,2,\ldots, \vspace*{-2mm}
\end{equation}
we can write \eqref{eq:I} as\vspace*{-5mm}

{{\small
\begin{eqnarray}
\mathcal{I} &=& \exp\left(-\alpha_{\mathrm{SR-SD}} \frac{\gamma(P_{\mathrm{PT}}x + N_0)}{P_{\mathrm{SR}}}\right)\nonumber \\
&& \times \!\sum_{k=0}^{m_{\mathrm{SR-SD}}-1}\! \dfrac{1}{k!} {\left(\alpha_{\mathrm{SR-SD}} \frac{\gamma(P_{\mathrm{PT}}x + N_0)}{P_{\mathrm{SR}}}\right)^k}\!\!.
\label{eq:I1}\vspace*{-2mm}
\end{eqnarray}}}\vspace*{-4mm}

\noindent Now, let\vspace*{-2mm}
\begin{equation}
\!\!\mathcal{A} =1 -\mathrm{Pr}(\gamma_{\mathrm{SR}_{i}} \leq \gamma ) = \mathrm{Pr}\!\left(\!\frac{P_{\mathrm{ST}}|h_{\mathrm{ST}-\mathrm{SR}_{i}}|^2}{P_{\mathrm{PT}}|h_{\mathrm{PT}-\mathrm{SR}_{i}}|^2 + N_0}> \gamma \!\right).
\label{eq:AA}
\end{equation}
Using the procedure to derive \eqref{eq:power}, we can write \eqref{eq:AA} as\vspace*{-4mm}

{{\small
\begin{eqnarray}
\mathcal{A} &=&\frac{\alpha_{\mathrm{PT-SR}}^{m_{\mathrm{PT-SR}}}\exp\left(\frac{-\alpha_{\mathrm{ST-SR}}\gamma N_0}{P_{\mathrm{ST}}}\right)}{\Gamma(m_{\mathrm{PT-SR}})}\nonumber \\
&&\hspace*{-10mm}\times \sum_{k=0}^{m_{\mathrm{ST-SR}}-1}\left(\frac{\alpha_{\mathrm{ST-SR}}
\gamma N_0}{P_\mathrm{ST}}\right)^k\frac{1}{k!} \times \left(\sum_{t=0}^{k}{\binom{k}{t}}\left(\frac{P_{\mathrm{PT}}}{N_0}\right)^t \right.\nonumber \\ 
&&\left.\hspace*{-10mm}\times\frac{\Gamma(m_{\mathrm{PT-SR}} + t)}{\left(\frac{\alpha_{\mathrm{ST-SR}}\gamma P_{\mathrm{PT}}}{P_{\mathrm{ST}}} + \alpha_{\mathrm{PT-SR}}\right)^{m_{\mathrm{PT-SR}} + t}}\right).
\label{eq:A}
\end{eqnarray}}}\vspace*{-3mm}

\noindent Substituting \eqref{eq:I1} and \eqref{eq:A} in \eqref{eq:main} and using the multinominal theorem \cite{gradshteyn}, we get \vspace*{-3mm}

{{\small
\begin{eqnarray}
&&\hspace*{-7mm}\mathrm{Pr}\left(\gamma_{\mathrm{tot}}^{N} \leq \gamma \big|_{|h_{\mathrm{PT-SD}}|^2 = x}\right) \nonumber\\
&\!\!\!\!\hspace*{-3mm} =&\hspace*{-3mm} \!\!\!\! \sum_{r_0=0}^N \sum_{r_1=0}^{r_0} \ldots \sum_{r_{m_{\mathrm{SR-SD}}-1}=0}^{r_{m_{\mathrm{SR-SD}}-2}} {\binom{N}{r_0}} {\binom{r_0}{r_1}} \ldots {\binom{r_{m_{\mathrm{SR-SD}}-2}}{r_{m_{\mathrm{SR-SD}}-1}}}\nonumber \\
 &\!\!\!\!\hspace*{-3mm} \times & \hspace*{-3mm} \!\!\!\mathcal{A}^{N-r_{m_\mathrm{SR-SD}-1}} (-1)^{N+r_{m_\mathrm{SR-SD}-1}}  \nonumber \\
 &\!\!\!\!\hspace*{-3mm} \times&\hspace*{-3mm} \!\!\!\mathrm{exp}\!\left(\! -\frac{\alpha_{\mathrm{SR-SD}}\gamma N_0(N-r_{m_\mathrm{SR-SD}-1})}{P_{\mathrm{SR}}} \right)\left( \frac{\alpha_{\mathrm{SR-SD}}\gamma N_0}{P_{\mathrm{SR}}} \right)^{R_{m_\mathrm{SR-SD}}} \nonumber\\
 &\!\!\!\!\hspace*{-3mm}\times&\!\!\!\!\hspace*{-3mm} \prod_{k=1}^{m_\mathrm{SR-SD}-1} \left( \frac{1}{k!}\right)^{r_{k-1}- r_{k}} \left( 1 + \frac{x P_{\mathrm{PT}}}{N_0}\right)^{R_{m_\mathrm{SR-SD}}} \nonumber \\ 
&\!\!\!\!\hspace*{-3mm}\times&\hspace*{-3mm}\!\!\!  \mathrm{exp}\left( -\frac{\alpha_{\mathrm{SR-SD}}\gamma P_{\mathrm{PT}} x(N-r_{m_\mathrm{SR-SD}-1})}{P_{\mathrm{SR}}} \right),
\end{eqnarray}}}\vspace*{-3mm}

\noindent where $R_{m_\mathrm{SR-SD}} = \sum_{k=1}^{m_{\mathrm{SR-SD}}-1}k(r_{k-1}-r_k)$. In the following step, we use binomial expansion of $\left( 1 + \frac{x P_{\mathrm{PT}}}{N_0}\right)^{R_{m_\mathrm{SR-SD}}}$ and take expectation over $|h_{\mathrm{PT-SD}}|^2$. Then, we can write\vspace*{-4mm}

{{\small
\begin{eqnarray}
{\mathrm{P}}^{N}_{\mathrm{s, out}}(\gamma)\!\!\!\!\!\! &=&\!\!\!\!\!\! \sum_{r_0=0}^{N}\sum_{r_1=0}^{r_0}\!\dotsc\!\!\!\!\!\! \sum_{r_{m_{\mathrm{SR-SD}}-1}=0}^{r_{m_{\mathrm{SR-SD}}-2}}\!\!{\binom{N}{r_0}}\!\!{\binom{r_0}{r_1}}\!\dotsc\!{\binom{r_{m_{\mathrm{SR-SD}}-2}}{r_{m_{\mathrm{SR-SD}}-1}}}\nonumber\\
&&\hspace*{-15mm}\!\!\!\!\!\times \mathcal{A}^{N-r_{m_\mathrm{SR-SD}-1}}	(-1)^{N+r_{m_{\mathrm{SR-SD}}-1}}  \nonumber \\ 
&&\hspace*{-15mm}\!\!\!\!\!\times  \left[\mathrm{exp}\left( -\frac{\alpha_{\mathrm{SR-SD}}\gamma N_0(N-r_{m_\mathrm{SR-SD}-1})}{P_{\mathrm{SR}}} \right)\right. \nonumber\\
&&\left.\hspace*{-15mm}\!\!\!\!\!\times \left(\frac{\alpha_{\mathrm{SR-SD}}\gamma N_0}{P_{\mathrm{SR}}}\right)^{R_{m_\mathrm{SR-SD}}}  \prod_{k=1}^{m_\mathrm{SR-SD}-1} \left( \frac{1}{k!}\right)^{r_{k-1}- r_{k}} \right] \nonumber \\
&&\hspace*{-15mm} \!\!\!\!\!\times  \Bigg[\sum_{p =0}^{R}{\binom{R}{p}}\left(\frac{P_{\mathrm{PT}}}{N_0}\right)^p \nonumber\\
&&\hspace*{-15mm}\!\!\!\!\!\times \int_{x=0}^\infty x^p \mathrm{exp}\left( -\frac{\alpha_{\mathrm{SR-SD}}\gamma P_{\mathrm{PT}} x(N-r_{m_\mathrm{SR-SD}-1})}{P_{\mathrm{SR}}} \right) \nonumber \\ 
&&\hspace*{-15mm}\!\!\!\!\! \times \frac{\alpha_{\mathrm{PT-SD}}^{m_{\mathrm{PT-SD}}}}{(m_{\mathrm{PT-SD}}-1)!} x^{m_{\mathrm{PT-SD}}-1} \mathrm{exp} \left(-\alpha_{\mathrm{PT-SD}} x\right) \mathrm{d}x\Bigg].
\label{eq:at1}
\end{eqnarray}}}\vspace*{-3mm}

\noindent Solving the integration in \eqref{eq:at1}, we get the required closed-form expression for the secondary outage probability given by \eqref{eq:psout}.\vspace*{-1mm}

\bibliographystyle{ieeetr}
\bibliography{paper}
\end{document}